\documentclass[aps,prx,twocolumn,superscriptaddress,english]{revtex4-1}
\usepackage[T1]{fontenc}
\usepackage[latin9]{inputenc}
\usepackage{amsmath}
\usepackage{amssymb}
\usepackage{graphicx}
\usepackage{float}
\usepackage{xcolor}
\usepackage{amsthm}

\makeatletter

\pdfpageheight\paperheight
\pdfpagewidth\paperwidth

\PassOptionsToPackage{position=top}{subfig}

\makeatother

\usepackage{babel}
\begin{document}
	\title{Symmetric non-Hermitian skin effect with emergent nonlocal correspondence}
	\author{Zhi-Yuan Wang}
	\affiliation{International Center for Quantum Materials and School of Physics, Peking University, Beijing 100871, China}
	\affiliation{Hefei National Laboratory, Hefei 230088, China}
	\author{Jian-Song Hong}
	\affiliation{International Center for Quantum Materials and School of Physics, Peking University, Beijing 100871, China}
	\affiliation{Hefei National Laboratory, Hefei 230088, China}
	\author{Xiong-Jun Liu}
	\thanks{Corresponding author: xiongjunliu@pku.edu.cn}
	\affiliation{International Center for Quantum Materials and School of Physics, Peking University, Beijing 100871, China}
	\affiliation{Hefei National Laboratory, Hefei 230088, China}
	\affiliation{International Quantum Academy, Shenzhen 518048, China}	
	\affiliation{CAS Center for Excellence in Topological Quantum Computation, University of Chinese Academy of Sciences, Beijing 100190, China}

\begin{abstract}
The non-Hermitian skin effect (NHSE) refers to that an extensive number of eigenstates of a non-Hermitian system are localized in open boundaries. Here we predict a universal phenomenon that the local particle-hole(-like) symmetry (PHS) leads to nonlocal pairs of skin modes distributed on different boundaries, manifesting a novel nonlocalization of the local PHS, which is unique to non-Hermitian systems. We develop a generic theory for the emergent nonlocal symmetry-protected NHSE by connecting the non-Hermitian system to an extended Hermitian Hamiltonian in a quadruplicate Hilbert space, which maps the skin modes to the topological zero modes, and the PHS to an emergent nonlocal symmetry in the perspective of many body physics. The predicted nonlocal NHSE is robust against perturbations. We propose optical Raman lattice models to observe the predicted phenomena in all physical dimensions, which are accessible with cold-atom experiments.
\end{abstract}
	
	\maketitle
	
	{\em Introduction.}--The non-Hermitian physics are ubiquitous in the systems effectively described by the non-Hermitian Hamiltonians, such as open quantum systems~\cite{Rotter2009, Cavity2010, Cavity2011}, quasiparticles in many body systems~\cite{Fu2017, Fu2018, Fu2019, Fu2020}, the acoustic and photonic systems with gain and loss~\cite{Ozawa2019, Photonic2019, Longhi2017, EP2021}, and so on. There are three important topics attracting theoretical interests: (i) the validity of non-Hermitian Hamiltonians in characterizing open systems~\cite{Bender1998, Bender2007, NHP2011}, (ii) the topological classification of the non-Hermitian lattice systems~\cite{Gongtopo2019, Borgnia2020, Kohomoto2011, Anomalous2016, Futopo2018, Wang2018, WangChern2018, Ueda2018, Satotopo2019, Ueda2020} or exception points (EPs)~\cite{EP2016, Rotter2016, EP2021, EP2022}, as the generalization of the well-known classification theory for Hermitian topological insulators or semimetals~\cite{Kane2010, Zhang2011, Ryu2016}, and (iii) the non-perturbative breakdown of the Bloch theorem due to the sensitivity to the boundary conditions~\cite{Wang2018, Wang2019, BBC2018, Ye2018, Murakami2019, Sato2019, Gong2019, Chiral2019, Transfer2019, biBBC2019, Sato2020, SatoHigh2020, Satoopen2021, Sato2023, Fang2020, Hu2020, Fang2022, Gong2020, Chen2020, Mirror2020, Lee2022, Singularity2020, Reciprocal2020, Stability2020, Floquet2021, Liew2021,Pseudo2022}. The key result of the last one, termed non-Hermitian skin effect (NHSE), is a phenomenon describing the localization of a volume-law number of eigenstates at the boundary under the open boundary condition (OBC). The skin modes refer to those localized states that are absent in Hermitian systems.

Being localized states on the boundary, the skin modes share similarities with but also have differences from topological boundary modes. 
A skin mode can be determined by the winding number of the complex spectrum under the periodic boundary condition (PBC)~\cite{Fang2020, Fang2022}, while such winding number is defined for each fixed mode, rather than on the whole bulk. The skin mode may also be interpreted by the topology of a related extended Hermitian Hamiltonian, which the non-Hermitian system is one-half of~\cite{Sato2019, Sato2020, Sato2023}, 
while the skin modes and topological boundary modes are different in nature. Further, while the existence of the NHSE could be characterized without symmetries, the effects of time-reversal symmetry and spatial symmetry on the NHSE are also recently investigated~\cite{Fang2022, Sato2019, Sato2020, SatoHigh2020, Chen2020, Mirror2020,  Reciprocal2020}, showing that the symmetries may enrich the features of the NHSE.

In this letter, we predict a universal phenomenon for non-Hermitian system that when the local particle-hole(-like) symmetry (PHS) is present, the non-Hermitian skin modes must appear in nonlocal pairs, localized in spatially separated open boundaries, and further propose realistic models for experimental observation. While the PHS is local, we show with a sophisticated proof that the PHS emerges as a nonlocal one on the skin modes, and transforms the skin mode in one boundary to that in another
(in the opposite boundary if under rectangular geometry). This result manifests a profound nonlocalization of the local PHS on the non-hermitian skin modes, 
which can be further interpreted with the many-body physics in the Hermitian counterparts. Moreover, the two skin mode on different boundaries and connected by the PHS are not degenerate, but have opposite eigenvalues. Thus the skin modes with emergent nonlocal correspondence are robust to the perturbations. The experimental realizations of the present study are proposed. 

\begin{figure}[tp]
 \centering
 \includegraphics[width=1.0\columnwidth]{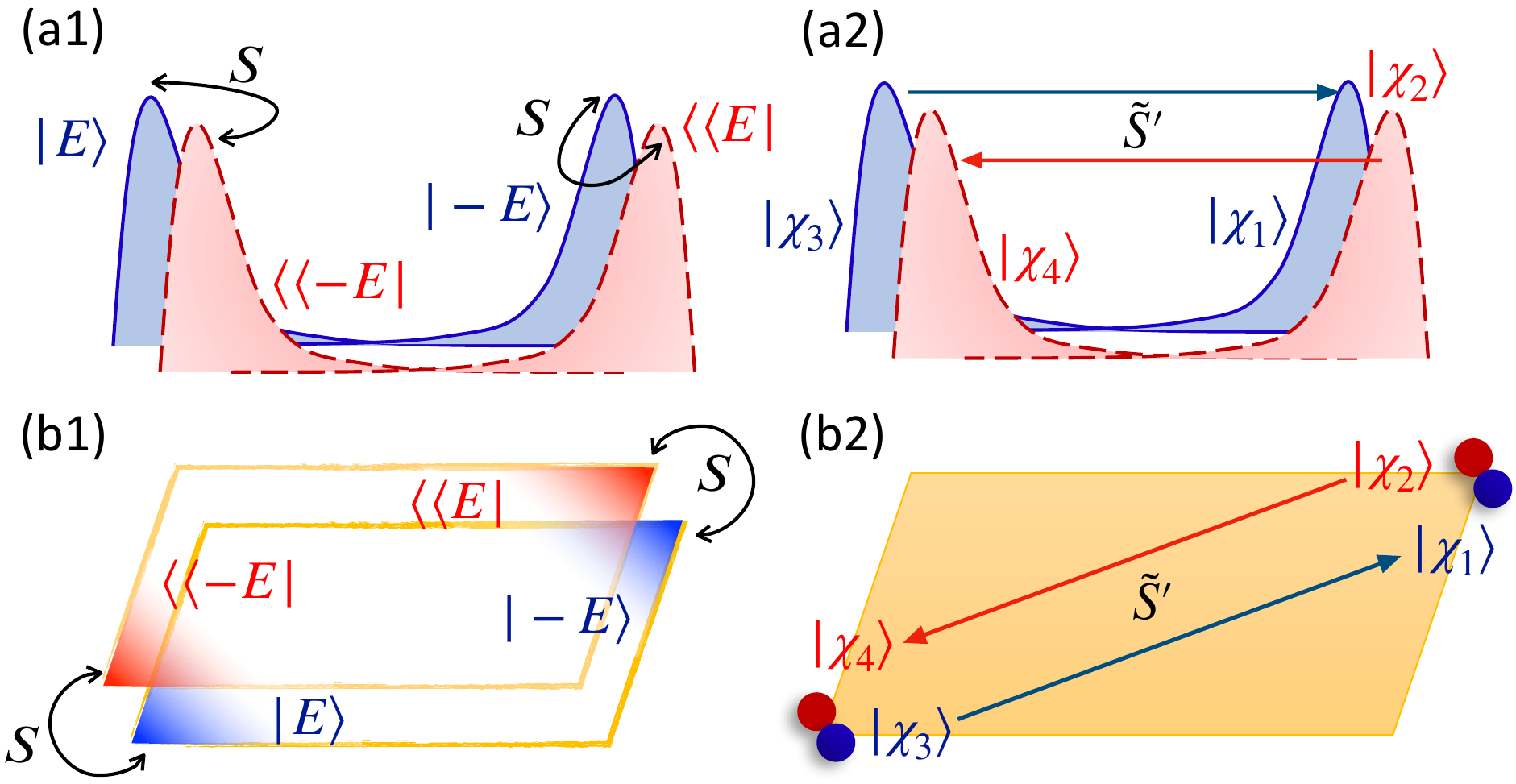}
 \caption{\label{Fig:mapping} The mapping between the skin modes of non-Hermitian Hamiltonian and the topological zero modes of the extended Hermitian Hamiltonian in the 1D and 2D cases. In (a1, b1) the blue (red) regions denote the right (left) eigenstates' wavefunction distributions of the skin modes. The right eigenvector $|E\rangle_{\rm skin}$ and the left eigenvector $\langle\langle-E|_{\rm skin}$ (the subscripts are omitted in the figures) are locally related by the PHS $S=UP$ and so do $\langle\langle-E|_{\rm skin}$ and $|-E\rangle_{\rm skin}$. By the construction of the extended Hamiltonian in Eq.~\eqref{eq.quadruple}, these right (left) eigenvectors are mapped to the topological zero modes $|\chi_{3}\rangle_{\rm topo}$ and $|\chi_{1}\rangle_{\rm topo}$ ($|\chi_{2}\rangle_{\rm topo}$ and $|\chi_{4}\rangle_{\rm topo}$) in (a2, b2). And the PHS $S$ is mapped to the emergent nonlocal symmetry $\tilde{S}'$ that relates the spatially separated zero modes when all negative energy bands of the Hermitian Hamiltonian are filled (denoted by the colored background in (b2)).}
\end{figure}

We develop a generic theory 
which is valid in arbitrary dimensions, 
without relying on the details of the Hamiltonian, applicable to amorphous systems and fractals, 
nor on the generalized Brillouin zone (GBZ) theory which is not reliable in high dimensions~\cite{Wang2018, Murakami2019, Fang2020, Hu2020}. Fig.~\ref{Fig:mapping} outlines the basic idea of the theory. We construct an extended Hermitian topological Hamiltonian in a quadruplicate Hilbert space, which the generic non-Hermitian system under consideration is one-quarter of. The skin modes and the PHS of the non-Hermitian Hamiltonian are mapped to topological zero modes and a new symmetry of the extended Hermitian system, respectively.
In the many-body physics level, one shall show that the local symmetry emerges as a nonlocal one if projected onto the topological zero modes on boundaries. This further transforms the skin modes localized on different boundaries when mapping back to the non-Hermitian system. 

{\em The generic non-Hermitian system and the extended Hamiltonian.}--We start with a generic non-Hermitian system whose first-quantized Hamiltonian $H$ respects the PHS $S=UP$ as~\cite{SRyu2010NJP}
\begin{equation}
UH^TU^{-1}=-H,\ UU^*=1,
\end{equation}
where $U$ is unitary and local, and $P$ is the transpose operation. If $|E\rangle_{\rm skin}$ is a right eigenstate with eigenvalue $E$, i.e. $H|E\rangle_{\rm skin}=E|E\rangle_{\rm skin}$, then there is a locally related left eigenstate $\langle\langle-E|_{\rm skin}=(U^{\dagger}|E\rangle_{\rm skin})^{T}$ with eigenvalue $-E$, i.e. $\langle\langle-E|_{\rm skin}H=-E\langle\langle-E|_{\rm skin}$. We shall prove that $|E\rangle_{\rm skin}$ and $|-E\rangle_{\rm skin}$ (the right eigenvector dual of $\langle\langle -E|_{\rm skin}$) are always localized at opposite boundaries, rendering an emergent nonlocal correspondence of the skin modes originated from the local PHS. This prediction is universal and is nontrivial, since the microscopic Hamiltonian does not have any nonlocal symmetries to relate the spatially separated skin modes. 

Our main idea to prove the emergent nonlocal correspondence is based on an extended Hermitian Hamiltonian we construct in this work that 
\begin{equation}\label{eq.quadruple}
\tilde{H}_{E}=\left(\begin{array}{cccc}
 &  &  & H^\dagger+E^*\\
 &  & H-E\\
 & H^{\dagger}-E^{*}\\
H+E
\end{array}\right).
\end{equation}
Note that $E \neq 0$ for the present consideration, since the degeneracy of $|E=0\rangle_{\rm skin}$ and $\langle\langle-E=0|_{\rm skin}$ will force them to be extended in general~\cite{SM}. Before proceeding to the rigorous proof of the emergent nonlocal correspondence, we point out several important properties of the extended Hamiltonian. We start with three properties associated with the first-quantized Hamiltonian $\tilde{H}_{E}$. Firstly, from the PHS of the original Hamiltonian $H$, we obtain the PHS of the extended Hamiltonian, $\tilde{S}=\tilde{U}\tilde{P}$, satisfying $\tilde{U}\tilde{H}_{E}^T\tilde{U}^{-1}=-\tilde{H}_{E}$, where $\tilde{P}$ is the transpose operation on the extended Hamiltonian and
\begin{equation}
\tilde{U}=\begin{pmatrix}
&U&&\\
U&&&\\
&&&-U\\
&&-U&
\end{pmatrix}.
\end{equation}
Further, the Hermitian Hamiltonian $\tilde{H}_{E}$ has four topological zero modes mapped from the skin modes $|E\rangle_{\rm skin}$ and $|-E\rangle_{\rm skin}$ that
\begin{eqnarray}\label{eq.zeromodes}
\begin{split}
|\chi_{3}\rangle_{\rm topo} &= \begin{pmatrix}0,0,|E\rangle_{\rm skin},0\end{pmatrix}^T,\\
|\chi_{1}\rangle_{\rm topo} &= \begin{pmatrix}|-E\rangle_{\rm skin},0,0,0\end{pmatrix}^T,
\end{split}
\end{eqnarray}
and $|\chi_{2}\rangle_{\rm topo}=\tilde{U}|\chi_{1}\rangle_{\rm topo}^*$, $|\chi_{4}\rangle_{\rm topo}=-\tilde{U}|\chi_{3}\rangle_{\rm topo}^*$. Finally, the extended Hermitian Hamiltonian can be decomposed into two copies with a proper unitary transformation $\tilde{H}_{E}\rightarrow H_{+E}\oplus H_{-E}$, where $H_{\pm E}$ are the two decoupled blocks~\cite{SM} respecting the artificial chiral symmetry $\Gamma_{a(b)}H_{+E(-E)}\Gamma_{a(b)}^\dagger=-H_{+E(-E)}$, with $\Gamma_{a/b}={\rm diag}\{1,1,\mp1,\pm1\}$ in the basis before decomposition. Note that $\Gamma_{a(b)}$ are {\em not} symmetries of the first-quantized Hamiltonian $\tilde{H}_{E}$. 
However, from $\Gamma_{a(b)}$ we can define symmetries of $\tilde{H}_{E}$ in the second-quantization: $\hat{\Gamma}_{a(b)}=\Gamma_{a(b)}\hat{K}_{a(b)}\hat{P}_{a(b)}$, with the complex conjugation $\hat{K}_{a(b)}$ and $\hat{P}_{a(b)}=\prod_{js}(c_{js,a(b)}^{\dagger}+c_{js,a(b)})$ being the particle-hole transformation~\cite{Zirnbauer2020} on the $H_{+E}(H_{-E})$, 
where $j$ is position index and $s$ denotes the remaining degree of freedom [e.g. the (pseudo-)spin]. Combining $\tilde{S}$ and $\hat{\Gamma}_{a}$ we reach then an important symmetry $\tilde{S}'=\hat{\Gamma}_{a}\tilde{S}$ of $\tilde{H}_{E}$ in second-quantization, which is essential to our proof and relates $|\chi_3\rangle_{\rm topo}$ and $|\chi_1\rangle_{\rm topo}$ (and thus $|E\rangle_{\rm skin}$ and $|-E\rangle_{\rm skin}$) nonlocally. Unlike $\hat{\Gamma}_{a}$, we note that $\tilde{S}'$ is a symmetry directly mapped from the PHS of the original Hamiltonian $H$. Bearing these features in mind, we are now ready to establish the generic nonlocal correspondence between these topological zero modes and that between the non-Hermitian skin modes.

{\em Emergent nonlocal correspondence of the skin modes.}--Now we turn to the rigorous proof of our key result, which is organized in three basic steps. Firstly, we determine the localization properties of the topological zero modes and skin modes. For convenience we consider first the one-dimensional (1D) case, in which the bulk spectrum of $\tilde{H}_{E}$ is fully gapped~\cite{SM} and the topological zero modes are protected by the chiral symmetries $\Gamma_{a(b)}$ of the decoupled blocks. Since $U$ (or $\tilde{U}$) is a local operation, $|\chi_{1,2}\rangle_{\rm topo}$ must localize in the same end and so do $|\chi_{3,4}\rangle_{\rm topo}$. The topological zero modes $|\chi_{2}\rangle_{\rm topo}$ and $|\chi_{3}\rangle_{\rm topo}$ of the Hamiltonian $H_{+E}$ have opposite chiralities of $\Gamma_a$. Thus they must be localized in the opposite ends (i.e. left-hand and right-hand boundaries)~\cite{Wu2013, Liu2013, Bernevig2014, Zhang2021, Sato2023}, similar for $|\chi_{1,4}\rangle_{\rm topo}$. These results naturally hold for $d$-dimensional systems ($d>1$), but the difference is that the bulk of the extended Hermitian system is gapless if $|\pm E\rangle_{\rm skin}$ are first-order skin modes~\cite{Fang2022,SM} and is usually fully gapped if $|\pm E\rangle_{\rm skin}$ are $d$th-order skin modes~\cite{Gong2019,Sato2020,SatoHigh2020} (see a detailed discussion for different scenarios in Supplementary Material~\cite{SM}), manifesting a novel mapping between the non-Hermitian skin modes and zero corner modes of higher-order topological semimetals~\cite{Gong2018, Roy2019, Bound2020} or insulators~\cite{Quadrupole2017, Gong2018, Chiral2022} (see also Fig.~\ref{Fig:mapping}). In general, we can conclude that $|E\rangle_{\rm skin}$ and $\langle\langle-E|_{\rm skin}$ localize in one boundary, while $|-E\rangle_{\rm skin}$ and $\langle\langle E|_{\rm skin}$ in a different boundary.

Secondly, we show the emergent nonlocal correspondence between the topological zero modes given by the symmetries $\tilde{S}'=\hat{\Gamma}_a\tilde{S}$ mapped from the PHS of the original non-Hermitian system. 
To show the emergent nonlocal nature of $\tilde{S}'$, we examine its action on the many-body state $\gamma_3^{\dagger}|\Phi_{\rm{bulk}}\rangle$, with $\gamma_i^{\dagger}$ being the creation operator of the topological zero mode $|\chi_i\rangle_{\rm topo}$ and $|\Phi_{\rm{bulk}}\rangle=\prod_{E'<0}c_{E',a}^{\dagger}\prod_{E''<0}c_{E'',b}^{\dagger}|vac\rangle$ characterizing all the bulk states with negative energies. The two products denote the creation of the states for $H_{+E}$ and $H_{-E}$, respectively. With some algebra we can show that~\cite{SM}
\begin{equation}\label{manybodyrelation}
\tilde{S}' \gamma_3^{\dagger}|\Phi_{\rm{bulk}}\rangle = \gamma_1^{\dagger}|\Phi_{\rm{bulk}}\rangle.
\end{equation}
The key feature is that $\tilde{S}'$ leaves the many-body bulk state invariant, while transforms the zero modes. The underlying physics is that the symmetry connects the eigenmodes of opposite energies and thus operates differently on the bulk and zero boundary modes~\cite{SM}. This renders a nonlocal correspondence between the two zero boundary modes. To characterize the relation directly, we project the symmetry onto the boundary as $S_{\rm proj}=\Pi\tilde{S}'\Pi$, with $\Pi$ denoting projection of the many-body state onto the subspace of boundary zero modes. It follows then 
\begin{equation}\label{eq:correspondence1}
S_{\rm proj}|\chi_3\rangle_{\rm topo}=|\chi_1\rangle_{\rm topo}.
\end{equation}

	\begin{figure}[tp]
	\centering
		\includegraphics[width=1.0\columnwidth]{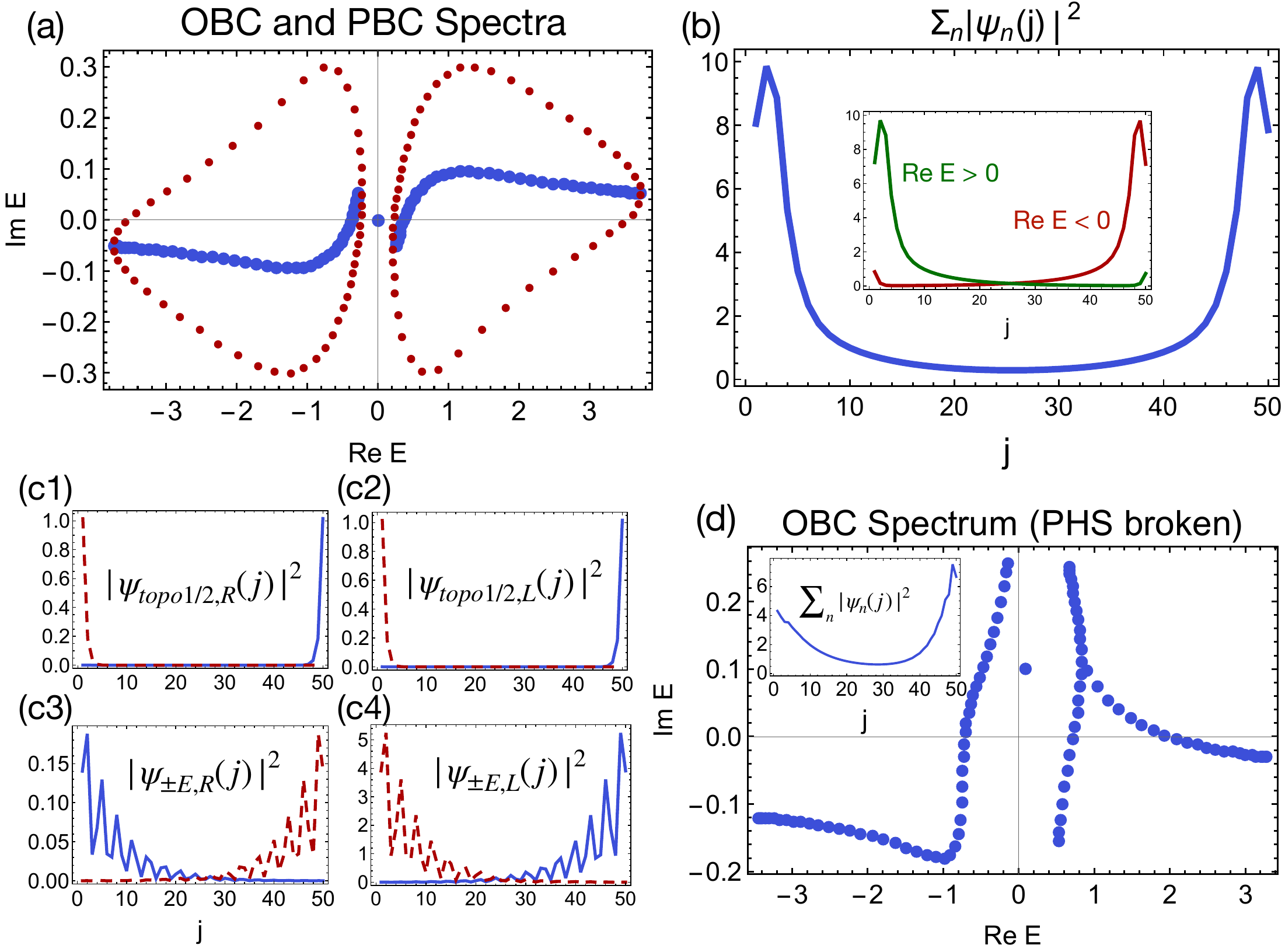}
		\caption{\label{Fig.1Dmain} The spectra, symmetric NHSE, localized states and the robustness of skin modes of the 1D model in Eq.~\eqref{eq.1Dmodelmain}. (a) The OBC (blue, larger dots) and PBC (red, smaller dots) spectra of the model in Eq.~\eqref{eq.1Dmodelmain} with $t_0=t_{so}=1.0, m_z=0.3, \gamma_\downarrow=0.6, K=2\pi/3$ and the size of the system $L=50$. (b) Density profile of eigenstates in the real space, summed over all eigenstates and spin degree of freedom, in which the inset plots the profile of the two bands' eigenstates respectively. (c1, c2) The wavefunction distribution of the pair of topological zero modes' right eigenvectors $\psi_{\rm topo1/2,R}$ (c1) and left eigenvectors $\psi_{\rm topo1/2,L}$ (c2). (c3, c4) The wavefunction distribution of a pair of skin modes' right eigenvectors $\psi_{\pm E,R}$ and left eigenvectors $\psi_{\pm E,L}$ with eigenenergy $\pm E=\pm(0.27-0.05\rm{I})$ (blue and red dashed respectively). (d) The OBC spectrum of the same model but with spin-dependent spin-conserved hopping coefficients $t_{0\uparrow}=0.5, t_{0\downarrow}=1.0$ breaking the PHS. The inset plots the density profile in this case, illustrating the skin modes distributed in opposite ends.}
		\end{figure}

Finally, we obtain the nonlocal correspondence for the skin modes of the non-Hermitian system. Note that the above zero modes are constructed from the skin modes according to Eq.~\eqref{eq.zeromodes}, and their corresponding wave functions are identical. 
Thus we know that the pair of skin modes $|E\rangle_{\rm skin}$ and $|-E\rangle_{\rm skin}$ are localized in opposite boundaries, and the correspondence Eq.~\eqref{eq:correspondence1} renders
\begin{equation}
S_{\rm proj}|E\rangle_{\rm skin}=|-E\rangle_{\rm skin}.
\end{equation}
Similarly, one can prove $S_{\rm proj}$ relates the other two zero modes $S_{\rm proj}|\chi_2\rangle_{\rm topo}=|\chi_4\rangle_{\rm topo}$ and so are the pair of skin modes for the left eigenvectors $\langle\langle E|_{\rm skin}S_{\rm proj}^\dag=\langle\langle-E|_{\rm skin}.$ Since the projective symmetry $S_{\rm proj}$ can always be constructed as long as the PHS exists for the non-Hermitian system, we have established the emergent nonlocal correspondence of the skin modes with local PHS.

A few remarks are worthwhile to provide. From the generic theory we know that the predicted symmetric NHSE does not necessitate any nonlocal symmetry but manifests a novel nonlocalization of the local PHS. Also, the proof and results are broadly applicable to non-Bloch systems including quasicrystals, amorphous systems, and fractals, because the above theory exploits only local symmetries and does not rely on spatial translational symmetry. See also numerical verification in Supplementary Material~\cite{SM}. Moreover, for completeness we discuss the topological zero modes with $E=0$, if existing for the non-Hermitian Hamiltonian, for which the above proof using extended Hermitian Hamiltonian no longer applies. We have shown that~\cite{SM}, for such zero modes, the right and left eigenstates $|E=0\rangle$ and $\langle\langle E=0|$ are localized in the same boundary [Fig.~\ref{Fig.1Dmain}(c1,c2)], in sharp contrast to the skin modes. Thus the skin modes and the topological boundary states are different in nature even they both exist as boundary states for a non-Hermitian system. Finally, in real experiment the symmetric NHSE can be observed even when the PHS is not perfectly satisfied and deviates to a certain extent [Fig.~\ref{Fig.1Dmain}(d)]. This robustness stems from the nondegeneracy of the two skin modes $|E\rangle_{\rm skin}$ and $|-E\rangle_{\rm skin}$, in sharp contrast to the symmetric NHSE related by time-reversal symmetry~\cite{Sato2020, Chen2020, Reciprocal2020} and the nonlocal spatial symmetry~\cite{SatoHigh2020, Mirror2020}, where the symmetry-related skin modes are degenerate and thus unstable against infinitesimal perturbations.

\begin{figure}[tp]
	\centering
		\includegraphics[width=1.0\columnwidth]{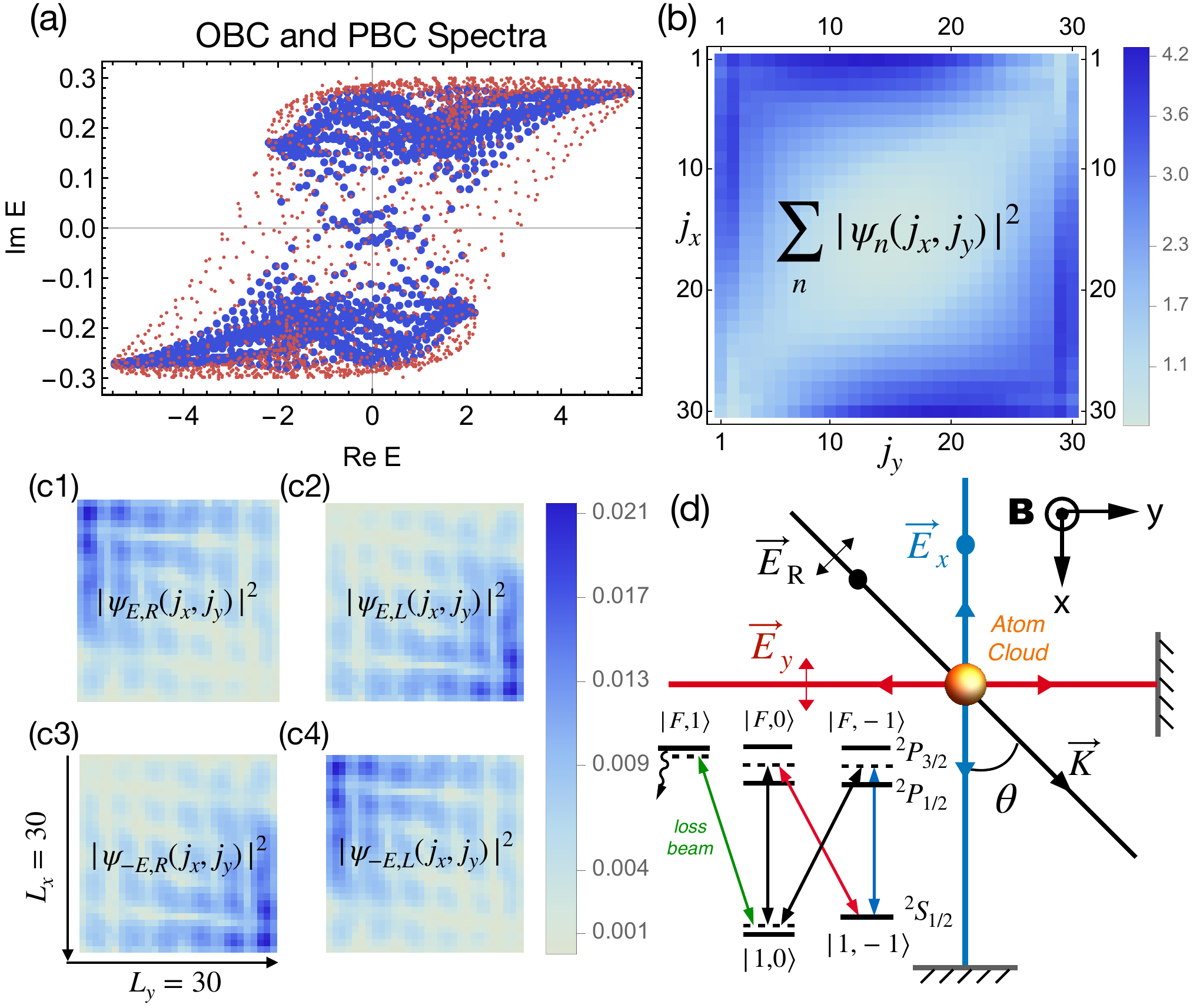}
		\caption{\label{Fig.2Dmain} The numerical results and experimental scheme of the 2D model in Eq.~\eqref{eq.2Dmodelmain}. (a) The OBC (blue, larger dots) and PBC (red, smaller dots) spectra of the model in Eq.~\eqref{eq.2Dmodelmain} with $t_0^x=t_0^y=1.0, t_{so}^x=-it_{so}^y=0.4,m_z=0.3,\gamma_\downarrow=0.6,\vec{K}=\pi/a(\cos\theta,\sin\theta),\theta=50^{\circ}$ and the size of the system $L_x=L_y=30$. (b) Denisty profile in the real space, summed over all eigenstates and the spin degree of freedom. (c1, c3) The real space distribution of  right eigenvectors $|\psi_{\pm E,R}|^2$ of a pair of skin modes with eigenenergies $\pm E=\mp(1.63+0.17\rm i)$ respectively, and (c2, c4) the distribution of corresponding left eigenvectors $|\psi_{\pm E,L}|^2$. (d) The scheme of Raman optical lattice experiment to realize the 2D model in Eq.~\eqref{eq.2Dmodelmain}. The green beam propagates along $z$ direction and is $\sigma^+$ polarized, inducing the effective damping of the ground state. The Raman beam's (black running wave) electric field $\vec{E}_{R}\propto ((\vec{e}_{x}-\vec{e}_{y})/\sqrt{2}+i\vec{e}_{z})e^{i\vec{K}\cdot\vec{j}}$ is circular polarized and the two linear components of it couple with the two optical lattice beams (red and blue) respectively, inducing the spin-flipped hoppings in the two directions.}
	\end{figure}

{\em Models and Experimental proposals}.--We propose now models in 1D, 2D, and 3D with the PHS to observe symmetric NHSE, which are experimentally feasible based on the Raman optical lattices as broadly studied in quantum simulation with ultracold atoms~\cite{Liu2013, Jo2018, Liu2018, LiuPan2018, Liu2020, LiuPan2021, Jo2022}. Denoting by $|js\rangle$ the single-particle state at $j$'s site with spin $s=(\uparrow,\downarrow)$, the Hamiltonian of the 1D lattice model is:
\begin{align}\label{eq.1Dmodelmain}
	H_{1D}	= &\sum\nolimits_{j}\Big[(m_{z}+i\gamma_{\downarrow}/2)\bigr(|j\uparrow\rangle\langle j\uparrow|-|j\downarrow\rangle\langle j\downarrow|\bigr)\\
	&-t_{0}\bigr(|j\uparrow\rangle\langle j+1\uparrow|-e^{-iK}|j\downarrow\rangle\langle j+1\downarrow|+h.c.\bigr)\nonumber\\
	&+t_{\rm so}\bigr(|j\downarrow\rangle\langle j+1\uparrow|-e^{iK}|j+1\downarrow\rangle\langle j\uparrow|+h.c.\bigr)\Big],\nonumber
\end{align}	
where the non-Hermiticity is given by $\gamma_{\downarrow}$ term, the Zeeman term $m_z$ and spin-conserved(flipped) hopping coefficient $t_{0(\rm so)}$ are controllable constants and $K$ is the projection of the Raman beam's wave vector on the axis of the optical lattice.  The PHS of this model is $S=UP=e^{-iKj}\sigma_x P$, with $P$ being the transpose operation. The centrosymmetric spectrum and the symmetric NHSE are shown in Fig.~\ref{Fig.1Dmain}(a,b,c1-c4), confirming the prediction from generic theory. The 1D model can be easily extended to the 2D one with Hamiltonian
\begin{align}\label{eq.2Dmodelmain}
	H_{2D}	&=\sum_{\vec{j}}\Big\{(m_{z}+i\gamma_{\downarrow}/2)\bigr(|\vec{j}\uparrow\rangle\langle\vec{j}\uparrow|-|\vec{j}\downarrow\rangle\langle \vec{j}\downarrow|\bigr)\\
	&-\sum_{k=x,y}\Big[t_0\bigr(|\vec{j}\uparrow\rangle\langle\vec{j}+\vec{e}_k\uparrow|-e^{-i\vec{K}\cdot \vec{e}_k}|\vec{j}\downarrow\rangle\langle\vec{j}+\vec{e}_k\downarrow|\bigr)\nonumber\\
	 &+t_{\rm so}^k\bigr(|\vec{j}\downarrow\rangle\langle\vec{j}+\vec{e}_k\uparrow|-e^{i\vec{K}\cdot \vec{e}_k}|\vec{j}+\vec{e}_k\downarrow\rangle\langle\vec{j}\uparrow|\bigr)+h.c.\Big]\Big\},\nonumber
	\end{align}
where $\vec{K}$ is the wave vector of the Raman beam in Fig.~\ref{Fig.2Dmain}(d) and $t_{\rm so}^k,k=(x,y)$ are the spin-flipped hopping coefficients in different directions. The corner skin effect shown in Fig.~\ref{Fig.2Dmain}(b) with most eigenstates being localized can be comprehended by the 1st-order NHSE under cylinder geometry~\cite{Fang2022, SM}. This model respects the PHS $S=UP=e^{-i\vec{K}\cdot\vec{j}}\sigma_x P$ and thus the corner skin modes appear in nonlocal pairs as illustrated by Fig.~\ref{Fig.2Dmain}(c1-c4). Without the damping term (i.e. $\gamma_{\downarrow}=0$) the above models are similar to the D class superconductors and are topologically characterized by $\mathbb{Z}_2$ (for 1D) and $\mathbb{Z}$ (for 2D) indices~\cite{Kane2010, Zhang2011, Ryu2016}. We have also studied the non-Hermitian 2D Dirac semimetal and 3D Weyl semimetal models, and confirmed the nonlocal symmetric NHSE. For simplicity we have put the details in Supplementary Material~\cite{SM}.

To see the experimental feasibility of the proposed models, we take the 2D model in Eq.~\eqref{eq.2Dmodelmain} as example. The experimental scheme illustrated by Fig.~\ref{Fig.2Dmain}(d) realizes a cold-atom $^{\rm{87}} \rm{Rb}$ system with Hamiltonian
=\begin{align}
	H_{2D}'	= &\int d^2\vec{r} \Big[ \sum_{s=\uparrow\downarrow} |\vec{r}s\rangle\bigr(-\frac{\hbar^2\vec{\nabla}^2}{2m}+V(\vec{r})+\frac{\delta}{2}(\sigma_z)_{ss}\bigr) \langle\vec{r}s|\nonumber\\
	-&i\gamma_{\downarrow}|\vec{r}\downarrow\rangle\langle\vec{r}\downarrow|+\bigr(M_R(\vec{r})|\vec{r}\uparrow\rangle\langle\vec{r}\downarrow|+h.c.\bigr)\Big],
	\label{eq.2Dcontinousmain}
\end{align}
with the spin up and down states defined as $|\uparrow\rangle=|1,-1\rangle$ and $|\downarrow\rangle=|1,0\rangle$ in the $F=1$ manifold. The non-Hermiticity comes from the damping of spin down state that could be effectively realized by the transition from spin down state to an excited atomic state with a short lifetime~\cite{Jo2022}, as induced by the $\sigma^+$ polarized loss beam, and the loss rate could be calibrated when the Raman beam is absent. The optical lattice potential $V(\vec{r})= V_{0x}\cos^2(k_0x)+V_{0y}\cos^2(k_0y)$ (with $V_{0x/y}\propto |\vec{E}_{x/y}^2|$) is driven by the standing waves along the 2D directions, and the Raman potential $M_R(\vec{r})=\bigr(M_{0x}\cos(k_0x)+M_{0y}\cos(k_0y)\bigr)e^{i\vec{K}\cdot\vec{r}}$ describes the two-photon processes (inset of Fig.~\ref{Fig.2Dmain}(d)) induced by the standing waves and running wave beam as $(M_{0x},M_{0y})\propto (E_{R}E_x,E_y E_{R})$ (see more details in Supplementary Material~\cite{SM}). Here the running wave beam $E_R$ is tilted to avoid the spatial symmetries coinciding with the boundary geometry~\cite{Fang2022}. Within the s-band tight-binding approximation, we can effectively reduce Eq.~\eqref{eq.2Dcontinousmain} to the lattice model in Eq.~\eqref{eq.2Dmodelmain} (up to an overall energy shift $-i\gamma_\downarrow/2$), 
with the coefficients given by~\cite{SM} $t_0=\int d^2\vec{r} \phi_{s}(\vec{r})[\frac{\hbar^2\vec{\nabla}^2}{2m}-V(\vec{r})]\phi_{s}(\vec{r}-a\vec{e}_k)$ and $t_{so}^k=\int d^2\vec{r} \phi_{s}(\vec{r})M_R(\vec{r})\phi_{s}(\vec{r}-a\vec{e}_k)$,
where $\phi_s$ is the s-band Wannier function. These coefficients are well controllable by tuning the beams independently. Note that while the symmetric NHSE exists as long as the damping term appears, the loss rate $\gamma_{\downarrow}$ should be neither too small or large compared with $t_{0,\rm so}$ since the system recovers the Hermiticity in the two extremes and the the skin effects are hard to observe~\cite{SM}. 
According to the previous studies~\cite{LiuPan2018, Jo2022} and the calculation with tight-binding approximation, here the appropriate parameters can be taken that $V_{0x/y}= 4,M_{0x/y}= 1,\delta=0.1,\gamma_\downarrow=0.1$ in the unit of the recoil energy $E_r=\frac{\hbar^2k_r^2}{2m}$, which give $t_{0}\approx 0.17,t_{so}^{x/y}\approx 0.07$ and $m_z=0.05$. The numerical results of the model with similar parameters given in Fig.~\ref{Fig.2Dmain} show the clear existence of NHSE in this regime. It is ready to know that the 1D model of Eq.~\eqref{eq.1Dmodelmain} can be realized by reducing the above lattice and Raman induced spin-orbit coupling to the 1D regime. 
Further, the 3D model can be obtained by introducing the damping term to the 3D optical Raman lattice~\cite{Liu2020,LiuPan2021}, which shall realize the 3D non-hermitian Weyl semimetal.

{\em Conclusions}.--We have uncovered a universal and broadly existing phenomenon of the emergence of nonlocal non-Hermitian skin effect (NHSE) when the local particle-hole(-like) symmetry (PHS) is present. For a non-Hermitian system with PHS and of arbitrary dimension, we showed that the skin modes always appear in pairs, with each being localized in different open boundaries. With a generic theory developed here, the non-Hermitian skin modes are mapped to the zero boundary or corner modes of gapped or gapless topological phases, through which an emergent nonlocal correspondence between the skin modes is established. 
This phenomenon is a novel manifestation of the nonlocalization of local PHS and is unique to the non-Hermitian systems. The universality of our prediction may open a new avenue to explore symmetry-protection features of NHSE and the related novel topological physics, and promote the investigation of NHSE in the high dimensions which is so far much less understood.
We have proposed lattice models for realizing the predicted symmetric NHSE, whose universality and robustness guarantee the high feasibility of the experimental observation.

\par We thank Gyu-Boong Jo for helpful discussions. This work was supported by National Key Research and Development Program of China (2021YFA1400900), the National Natural Science Foundation of China (Grants No. 11825401 and No. 12261160368), the Innovation Program for Quantum Science and Technology (Grant No. 2021ZD0302000), and the Strategic Priority Research Program of the Chinese Academy of Science (Grant No. XDB28000000).

	\onecolumngrid
	
	\renewcommand{\thesection}{S-\arabic{section}}
	\setcounter{section}{0}  
	\renewcommand{\theequation}{S\arabic{equation}}
	\setcounter{equation}{0}  
	\renewcommand{\thefigure}{S\arabic{figure}}
	\setcounter{figure}{0}  
	\newcommand{\diff}{\mathrm{d}}
	\newcommand{\biaoti}{\fontsize{12pt}{\baselineskip}\selectfont}

	\indent

	\begin{center}
\large \textbf{\large Supplementary Material:\\Symmetric non-Hermitian skin effect with emergent nonlocal correspondence}
\end{center}

In the \ref{sec.1} section we introduce in detail the extended Hermitian Hamiltonian method. In the \ref{sec.2} section we discuss the localization properties of the eigenstates. In the \ref{sec.3} section we derive the nonlocal correspondence of the skin modes due to the PHS. In the \ref{sec.4} section we propose the models in different dimensions exhibiting this symmetric NHSE and discuss the experimental feasibility in fermionic cold-atom systems.

\section{The extended Hermitian Hamiltonian formalism}\label{sec.1}
	In this section we provide details for the extended Hermitian Hamiltonian formalism of non-Hermitian system with PHS. The non-Hermitian single-particle Hamiltonian $H$ respects the PHS $S=UP$ as $U H^T U^{-1}=-H$, where $U$ is a unitary local operation satisfying $UU^*=1$ and $P$ is the transpose operation. The spectrum of this system is centrosymmetric due to the PHS, {\it i.e.}, if $|E\rangle$ is a right eigenstate of $H$ with eigenvalue $E$ as $H|E\rangle_{\rm skin}=E|E\rangle_{\rm skin}$, then $\langle\langle-E|_{\rm skin}=(U^{\dagger}|E\rangle_{\rm skin})^{T}$ is a left eigenstate with eigenvalue $-E$ as $\langle\langle-E|_{\rm skin}H=-E\langle\langle-E|_{\rm skin}$. To study the localization behavior, topological origin and relations of the skin modes, we map $H$ to the following extended Hermitian Hamiltonian
\begin{equation}
\tilde{H}_{E}=\left(\begin{array}{cccc}
 &  &  & H^\dagger+E^*\\
 &  & H-E\\
 & H^{\dagger}-E^{*}\\
H+E
\end{array}\right),
\end{equation}
which is defined on the quadruplicated Hilbert space and could be decomposed into two copies with a proper unitary transformation
\begin{equation}
\tilde{H}_{E}\rightarrow\left(\begin{array}{cc}
 & H-E\\
H^{\dagger}-E^{*}
\end{array}\right)\oplus\left(\begin{array}{cc}
 & H+E\\
H^\dagger+E^*
\end{array}\right)=H_{E}\oplus H_{-E},
\end{equation}
where $H_{\pm E}$ are the standard forms of doubled Hamiltonian. Note that here we focus on the $E\neq 0$ case since the degeneracy of $|E=0\rangle_{\rm skin}$ and $\langle\langle E=0|_{\rm skin}$ force them to be extended in general.

The extended Hamiltonian has both symmetries inherited from the non-Hermitian Hamiltonian and its own symmetries by construction. From the PHS of the original non-Hermitian Hamiltonian we obtain the PHS symmetry of the extended Hamiltonian, $\tilde{S}=\tilde{U}\tilde{P}$, satisfying $\tilde{U}\tilde{H}_{E}^T\tilde{U}^{-1}=-\tilde{H}_{E}$, where $\tilde{P}$ is the transpose operation on the extended Hamiltonian and
\begin{equation}
\tilde{U}=\begin{pmatrix}
&U&&\\
U&&&\\
&&&-U\\
&&-U&
\end{pmatrix}.
\end{equation}
Besides, $\tilde{H}_{E}$ respects the artificial chiral symmetry $\Gamma=\text{diag}(1,1,-1,-1)$ and thus the time-reversal symmetry $T=\Gamma\tilde{U}\hat{K}$ where $\hat{K}$ is the complex conjugation. Another important observation is that the two copies respects their own chiral symmetry $\Gamma_{a(b)}=\text{diag}(1,1,\mp 1,\pm 1)$ as $\Gamma_{a(b)}H_{E(-E)}\Gamma_{a(b)}^{-1}=-H_{E(-E)}$. Although $\Gamma_{a(b)}$ are not symmetries of the first-quantized Hamiltonian $\tilde{H}_E$, they become symmetries of the many-body Hamiltonian corresponding to $\tilde{H}_E$ after second quantization: $\hat{\Gamma}_{a(b)}=\Gamma_{a(b)}\hat{K}_{a(b)}\hat{P}_{a(b)}$, where $\hat{K}_{a(b)}$ and $\hat{P}_{a(b)}=\prod_{js}(c^\dagger_{js,a(b)}+c_{js,a(b)})$ are the complex conjugation and particle-hole conjugation on the $H_{E(-E)}$ degree of freedom (not affecting the other copy $H_{-E(E)}$), with $j$ and $s$ being the position and (pseudo-)spin indices.

An important property, whether the extended Hamiltonian is gapped or not, depends on the dimensions of the system. For 1D systems, if the eigenstate $|E\rangle_{\rm skin}$ with $E\neq0$ is a skin mode of the non-Hermitian Hamiltonian $H$, then $\tilde{H}_{E}$ possesses a gapped bulk spectrum, for which the proof is straightforward: $\det(\tilde{H}_{E}(\vec{k}))=|\det(H(\vec{k})-E)\det(H(\vec{k})+E)|^2\neq 0,\forall \vec{k}\in 1BZ$ because $E$ is excluded from the PBC spectrum of $H$, indicating that zero value is not included in the bulk continuum of $\tilde{H}_{E}$. As an aside, due to the chiral symmetry and the gapped bulk spectrum, $H_{\pm E}$ belongs to the AIII class classified by the $Z$ integer topological invariant $n_{\pm E}=\int\frac{dk}{2\pi i}\frac{\partial}{\partial k}\ln\det(H(k)\mp E)$, which is nothing but the winding number of the PBC spectral of $H$ around the reference point $\pm E$. In higher dimensions, the OBC spectrum of the non-Hermitian system is often covered by the PBC spectrum in the thermodynamic limit (see Fig.~\ref{Fig.2Dspec}(a1, b1) for example), and thus $\exists \vec{k}\in BZ,s.t.\det(\tilde{H}_{E}(\vec{k}))=0$, indicating the gapless bulk of $\tilde{H}_E$.

\section{Localization properties of the skin modes and topological zero modes}\label{sec.2}
In this section, we further map the skin modes to topological zero modes of the extended Hamiltonian and by this means discuss the localization properties of the skin modes. From the right and left eigenvectors of the pair of skin modes of $H$ with eigenenergy $\pm E$ we obtain four topological zero modes of $\tilde{H}_{E}$ under OBC
\begin{equation}\label{eq.zeromode}
|\chi_{1,2,3,4}\rangle_{\rm topo}=\left(\begin{array}{c}
|-E\rangle_{\rm skin}\\
\\
\\
\\
\end{array}\right)\left(\begin{array}{c}
\\
|E\rangle\rangle_{\rm skin}\\
\\
\\
\end{array}\right),\left(\begin{array}{c}
\\
\\
|E\rangle_{\rm skin}\\
\\
\end{array}\right),\left(\begin{array}{c}
\\
\\
\\
|-E\rangle\rangle_{\rm skin}
\end{array}\right).
\end{equation}
Since the eigenmodes of $H$ with eigeneneregy $E$ have an exact one-to-two correspondence to the zero modes of $\tilde{H}_E$, if no more symmetries of $H$ other than the PHS are assumed and thus both skin modes $|\pm E\rangle_{\rm skin}$ are non-degenerate, there are only four zero modes of $\tilde{H}_E$. These zero modes are related by the PHS $\tilde{S}=\tilde{U}\tilde{P}$ as
\begin{eqnarray}
\tilde{U}|\chi_{1}\rangle_{\rm topo}^*=|\chi_{2}\rangle_{\rm topo},&\ &\tilde{U}|\chi_{3}\rangle_{\rm topo}^*=-|\chi_{4}\rangle_{\rm topo},\nonumber\\
\tilde{U}|\chi_{2}\rangle_{\rm topo}^*=|\chi_{1}\rangle_{\rm topo},&\ &\tilde{U}|\chi_{4}\rangle_{\rm topo}^*=-|\chi_{3}\rangle_{\rm topo}.
\end{eqnarray}
The localization properties of the topological zero modes are determined by the symmetries. (i) Since $\tilde{U}$ is an local operation, $|\chi_{1,2}\rangle_{\rm topo}$ localize in the same boundary and so do $|\chi_{3,4}\rangle_{\rm topo}$. (ii)  In the 1D case, the topological zero modes $|\chi_{1}\rangle_{\rm topo}$ and $|\chi_{4}\rangle_{\rm topo}$ have opposite chiralities of $\Gamma_a$, thus localized in different ends. Therefore, $|\chi_{1}\rangle_{\rm topo}$ and $|\chi_{2}\rangle_{\rm topo}$ localize in the same end while $|\chi_{3}\rangle_{\rm topo}$ and $|\chi_{4}\rangle_{\rm topo}$ in the other end. For higher-dimensional systems, the only difference is that the bulk spectrum of $\tilde{H}_E$ is usually gapless, and the mapped zero modes are essentially the boundary (corner) modes of a higher-order topological semimetal. Without other symmetries' protection, there are no more zero modes and the eigenstates with opposite chiralities must localize separately. Then we can draw the conclusion holding in all dimensions: $|\chi_{1,2}\rangle_{\rm topo}$ localize in the same boundary while $|\chi_{3,4}\rangle_{\rm topo}$ in the opposite boundary.

For completeness, we discuss the case with $E=0$ where the extended Hamiltonian has more symmetries and the results above no longer holds. For the topological zero modes of the non-Hermitian Hamiltonian $H$ the analysis could be carried out without the help of the extended Hamiltonian. Applying the PHS  of $H$ on a topological zero mode of $H$, $|0\rangle$ and $\langle\langle 0|$, we obtain another pair of right and left eigenvectors $U|0\rangle\rangle^*$ and $\langle 0|^*U$. There are two possibilities: (i)$|0\rangle$ and $U|0\rangle\rangle^{*}$ are two different zero modes of $H$, which means $|0\rangle$ and $\langle0|U$ (as the left eigenvector of $U|0\rangle\rangle^{*})$ are Hermitian orthogonal, and (ii) $|0\rangle\propto U|0\rangle\rangle^{*}$ are the right eigenstates of the same zero mode. If (i) is true, $\langle0|U(|0\rangle^{*})=0$. But since in our case $UU^{*}=1$ (rather than $UU^{*}=-1$, in which case $\langle0|U(|0\rangle^{*})=\langle0|U^{T}(|0\rangle^{*})=-\langle0|U(|0\rangle^{*})=0)$, generally $\langle0|U|0\rangle^{*}$ should be nonzero if $|0\rangle$ is a topological zero mode. Thus $\langle0|U$ and $|0\rangle$ should be the left and right eigenvectors of the same topological zero mode of $H$ and are localized in the same end.

\section{Nonlocal correspondence of the skin modes}\label{sec.3}
	In this section we establish the nonlocal correspondence of the topological zero modes of the extended Hermitian Hamiltonian and then map this back to obtain the nonlocal correspondence of the non-Hermitian skin modes. We first show that the topological zero modes of $\tilde{H}_{E}$ localized in different boundaries are related by emergent nonlocal symmetry $\tilde{S}'=\hat{\Gamma}_a\tilde{S}$, which is the symmetry of the second-quantized Hamiltonian of $\tilde{H}_E$ and has no corresponding symmetry in the first-quantized form. The nonlocal nature of $\tilde{S}'$ is manifested by its action on the many-body state $\gamma^\dagger_{3}|\Phi_{\rm bulk}\rangle$ with $|\Phi_{\rm bulk}\rangle=\prod_{E'<0}c_{E',a}^{\dagger}\prod_{E''<0}c_{E'',b}^{\dagger}|vac\rangle$, where $\gamma_i^\dagger$ creates the topological zero mode $|\chi_i\rangle_{\rm topo}$ ($|\chi_3\rangle_{\rm topo}=(0,0,|E\rangle_{\rm skin},0)^T$) and the two products create all eigenstates with negative eigenenergies of $H_{E}$ and $H_{-E}$ respectively. By definition of $\hat{\Gamma}_{a}=\Gamma_a\hat{K}_a\prod_{js}(c^\dag_{js,a}+c_{js,a})$, we have
\begin{equation}
\hat{\Gamma}_{a}\gamma_{2,3}^{\dagger}\hat{\Gamma}_{a}^{-1}=\gamma_{2,3},\ \hat{\Gamma}_{a}c_{E',a}^{\dagger}\hat{\Gamma}_{a}^{-1}=c_{-E',a}.
\end{equation}
$\hat{\Gamma}_{a}$ doesn't act on the other block $H_{-E}$ and creates all the eigenstates of $H_{E}$ from the vacuum as $\hat{\Gamma}_{a}|vac\rangle=\gamma_2^\dagger\gamma^\dagger_3\prod_{E'}c_{E',a}^{\dagger}|vac\rangle$. The PHS $\tilde{S}$ acts on the eigen-operators as
\begin{equation}
\tilde{S}\gamma_{1,2,3,4}^\dag\tilde{S}^{-1}=\gamma_{2,1,4,3},\tilde{S}c^\dagger_{E',a(b)}\tilde{S}^{-1}=c_{-E',b(a)},
\end{equation}
and creates all the eigenstates of $\tilde{H}_E$ from the vacuum as $\tilde{S}|vac\rangle=\gamma_1^\dagger\gamma_2^\dagger\gamma_3^\dagger\gamma^\dagger_4\prod_{E'}c_{E',a}^{\dagger}c_{E',b}^{\dagger}|vac\rangle$.
Collecting the facts above,
\begin{align}
\tilde{S}' \gamma_3^{\dagger}|\Phi_{\rm bulk}\rangle &=\hat{\Gamma}_{a}(\tilde{S}\gamma_3^\dagger\tilde{S}^{-1})\prod_{E'<0}(\tilde{S}c_{E',a}^{\dagger}\tilde{S}^{-1})(\tilde{S}c_{E',b}^{\dagger}\tilde{S}^{-1})\tilde{S}|vac\rangle\nonumber\\
&=\hat{\Gamma}_{a}\gamma_4\prod_{E'>0}c_{E',b}c_{E',a}\gamma_1^\dagger\gamma_2^\dagger\gamma_3^\dagger\gamma_4^\dagger\prod_{E''}c_{E'',a}^{\dagger}c_{E'',b}^{\dagger}|vac\rangle\nonumber\\
&=\hat{\Gamma}_{a}\gamma_1^{\dagger}\gamma_2^{\dagger}\gamma_3^{\dagger}\prod_{E'<0}c_{E',a}^{\dagger}c_{E',b}^{\dagger}|vac\rangle\nonumber\\
&=\gamma_1^{\dagger}\gamma_2\gamma_3\prod_{E'<0}c_{-E',a}^{\dagger}c_{E',b}^{\dagger}\gamma_2^\dag\gamma_3^\dag\prod_{E''}c_{E',a}^{\dagger}|vac\rangle\nonumber\\
&=\gamma_1^{\dagger}|\Phi_{\rm bulk}\rangle.
\end{align}
Noting that $\tilde{S}'$ leaves the bulk states invariant and to obtain the direct relation between the zero modes, it's natural to project the symmetry onto the boundary $S_{\rm proj}=\Pi\tilde{S}'\Pi$, where $\Pi$ is the projection operator onto the subspace of boundary modes. In this way it's clear that $|\chi_1\rangle_{\rm topo}$ and $|\chi_3\rangle_{\rm topo}$ are nonlocally related by
\begin{equation}
S_{\rm proj}|\chi_3\rangle_{\rm topo}=|\chi_1\rangle_{\rm topo}.
\end{equation}
Finally, we map this nonlocal correspondence back to the non-Hermitian system and uncover the nonlocal correspondence of the skin modes. Remembering that the wavefunction of a skin mode is identical to that of the corresponding topological zero mode by Eq.~\eqref{eq.zeromode}, we know that the pair of skin modes $|E\rangle_{\rm skin}$ and $|-E\rangle_{\rm skin}$ are localized at opposite positions and are related to each other by
\begin{equation}
S_{\rm proj}|E\rangle_{\rm skin}=|-E\rangle_{\rm skin}.
\end{equation}
Similarily, $S_{\rm proj}$ relates the other two zero modes directly by $S_{\rm proj}|\chi_2\rangle_{\rm topo}=|\chi_4\rangle_{\rm topo}$ and so are the pair of skin modes' left eigenvectors as $\langle\langle E|_{\rm skin}S_{\rm proj}^\dag=\langle\langle -E|_{\rm skin}$. Since $S_{\rm proj}$ originates from the PHS of $H$, we have established the emergent nonlocal correspondence of the skin modes of non-Hermitian systems with local PHS. As a remark, since our whole proof just exploit the basic properties of the local symmetries, this should be an universal result for non-Hermitian systems with PHS, including the quasicrystals, amorphous systems and fractals. For example, one can see this symmetric NHSE in a quasiperiodic system as shown in Fig.~\ref{Fig.1Dspec}(b3). And the system if being Hermitian, {\em i.e.} the loss rate $\gamma_\downarrow=0$,  is in the localized phase and the total wavefunction distribution is uniform in real space.

	\section{Optical Lattice models with particle-hole symmetry in 1D, 2D and 3D}\label{sec.4}
	In this section we propose the non-Hermitian lattice models with the PHS in different dimensions that are readily realizable in optical Raman lattice experiments. For each model we show numerically the spectrum and the wavefunction distribution in real space and give the experimental scheme in Raman optical lattice.

	\subsection{1D non-Hermitian model with PHS}
\begin{figure}[h]
	\centering
		\includegraphics[width=1.0\columnwidth]{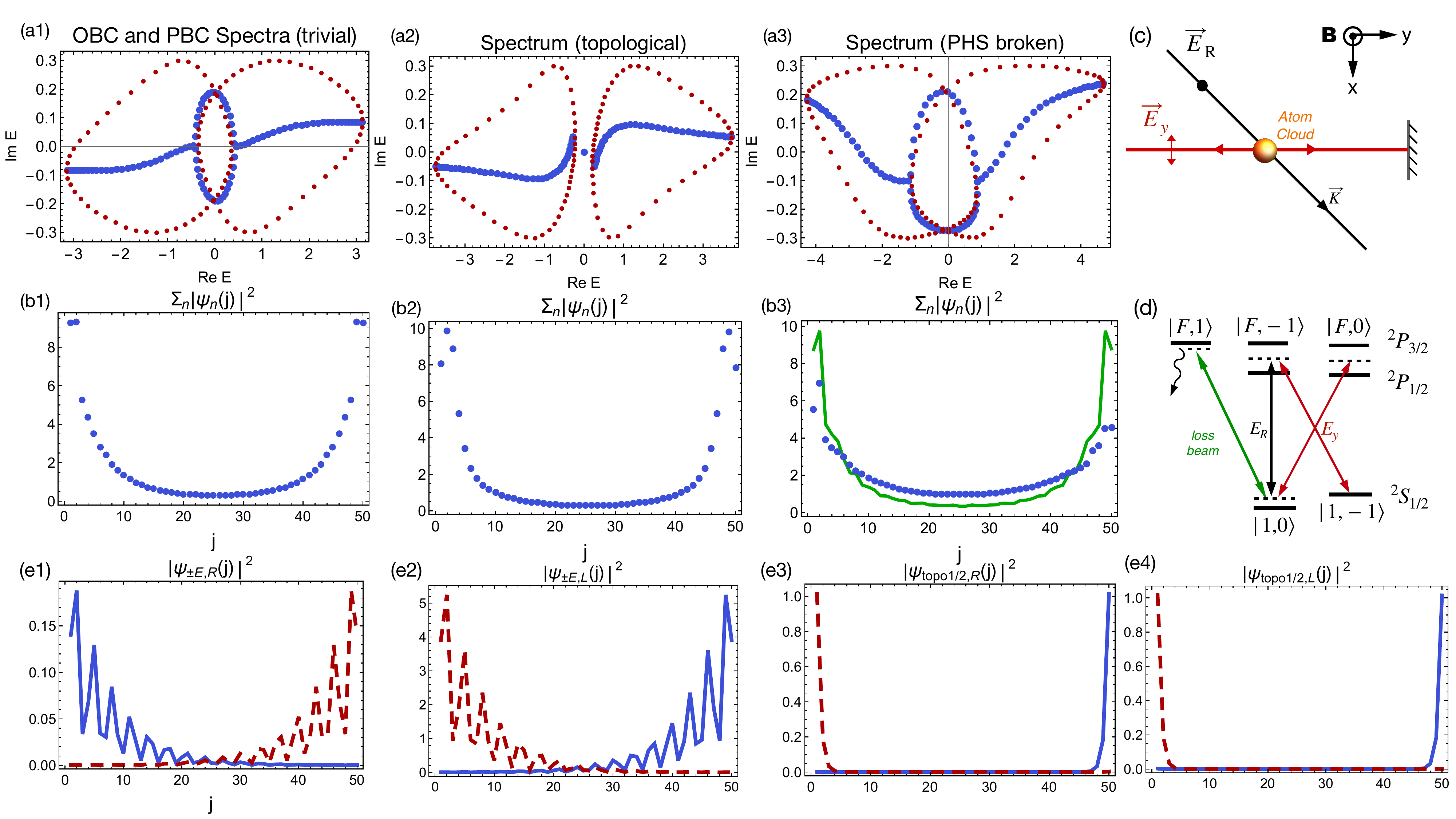}
		\caption{\label{Fig.1Dspec}Spectrum, wavefunction distributions and the experiment scheme of the 1D model. (a1, a2) The OBC (blue) and PBC (red) spectra of the model in Eq.~\eqref{eq.1Dmodel} with $t_0=1,m_z=0.3,\gamma_\downarrow=0.6,K=2\pi/3$ and $t_{so}=0.7,1.0$ respectively, corresponding to the trivial and topological cases. (a3) The OBC spectrum when $t_{so}=0.7$, $t_\downarrow=2t_\uparrow=2$. (b1, b2) The real space wavefunction distribution (summed over all right eigenstates and spin degree of freedom) for the trivial and topological cases. (b3) The wavefunction distribution when the symmetry is broken (blue dots) corresponding to (a3), and when the Zeeman term in Eq.~\eqref{eq.1Dmodel} is in a quasiperiodic form $(m_z\cos(2\pi j\alpha)+\rm I\gamma_\downarrow/2)\sigma_z$ (green line), where $m_z=0.3$ and $\alpha=(\sqrt{5}-1)/2$. (c) The scheme of Raman optical lattice experiment to realize the 1D model in Eq.\eqref{eq.1Dmodel}. The $^{\rm{87}}\rm{Rb}$ atoms are confined in the 1D optical lattice along $y$ direction formed by the standing wave (red beam) and the spin states are defined to be the atomic states $|\uparrow\rangle=|1,-1\rangle,|\downarrow\rangle=|1,0\rangle$ of the $F=1$ manifold in the hyperfine structure. (d) All relevant transitions for the atoms, including the two photon process driven by the Raman beam (black running wave in (c)) and the standing wave inducing spin-flipped couplings (SOC). The loss beam (green) is applied to drive the transition between the spin down state and some excited state with a short lifetime, giving the damping of the spin down state effectively. (e1-e4) The wavefunction distributions of the right and left eigenvectors of a pair of skin modes $\psi_{\pm E,R/L}$ with eigenenergy $\pm E=\pm(0.27-0.05\rm{I})$ (denoted by red and blue respectively) and the topological zero modes $\psi_{\rm topo1/2,R/L}$ of the model in Eq.~\eqref{eq.1Dmodel} when $t_{so}=1.0$.}
	\end{figure}

	Based on the researches on the 1D Raman optical lattices~\cite{Liu2013SM, Jo2018SM}, here we propose a 1D non-Hermitian lattice model with the PHS which exhibits the nonlocal symmetric skin effect as we predicted:	
\begin{align}
	H_{1D}	= &	\sum_{j}(m_{z}+i\gamma_{\downarrow}/2)\bigr(|j\uparrow\rangle\langle j\uparrow|-|j\downarrow\rangle\langle j\downarrow|\bigr)-t_{0}\bigr(|j\uparrow\rangle\langle j+1,\uparrow|-e^{-iK}|j\downarrow\rangle\langle j+1\downarrow|+h.c.\bigr)\nonumber\\
	&+t_{so}\bigr(|j\downarrow\rangle\langle j+1\uparrow|-e^{iK}|j+1\downarrow\rangle\langle j\uparrow|+h.c.\bigr), \label{eq.1Dmodel}
\end{align}	
where the non-Hermiticity is given by $\gamma_\downarrow$ term, the Zeeman term $m_z$ and spin-conserved (flip) hopping coefficient $t_{0(so)}$ are controllable constants, and $K$ is the projection of the Raman beam's wave vector on the axis of the optical lattice. The PHS is $S=UP=e^{-iKj}\sigma_x P$ where the Pauli matrix acts on the spin degree of freedom at the $j$'s site, and $P$ is the transpose operation. As we can see in Fig.~\ref{Fig.1Dspec}(b1-b3), the total wavefunction distribution of this model is symmetric because the skin modes related by the PHS show up in different ends symmetrically. This also agrees with the fact that the spectral winding numbers around the interior points of the band are $\pm1$ for the two PBC bands on the right and left on the complex plane respectively in Fig.~\ref{Fig.1Dspec}(a1, a2). Moreover, the symmetric skin effect is robust against perturbations whether breaking the PHS or not, because the PHS itself doesn't bring degeneracy in general and thus the skin modes in different ends are not easily mixed by the perturbations. We can see this robustness in Fig.~\ref{Fig.1Dspec}(a3, b3). Besides, a 1D non-Hermitian system with a point gap and respecting the PHS is topologically classified by a $Z_2$ index~\cite{Satotopo2019SM} and the model here could be either trivial or topological as shown in Fig.~\ref{Fig.1Dspec}(a1, a2) corresponding to the index $\nu=0,1$ respectively. In the topological case there is a pair of topological zero modes similar to the majorana zero modes in topological superconductors, whose right eigenvectors and left eigenvectors are localized in the same end just as predicted, while a skin mode's right and left eigenvectors are always localized in different ends as shown in Fig.~\ref{Fig.1Dspec}(e1-e4).

The experiment scheme illustrated by Fig.~\ref{Fig.1Dspec}(c, d) prepares a $^{\rm{87}} \rm{Rb}$ cold-atom system described by
 \begin{align}
	H_{1D}'	= &\int dy \Big[ \sum_{s=\uparrow,\downarrow} |ys\rangle\bigr(-\frac{\hbar^2\partial_y^2}{2m}+V(y)+\frac{\delta}{2}(\sigma_z)_{ss})\bigr)\langle ys|\nonumber\\
	&-i\gamma_{\downarrow}|y\downarrow\rangle\langle y\downarrow|+\bigr(M_R(y)|y\uparrow\rangle\langle y\downarrow|+h.c.\bigr)\Big].
\end{align}
where $\delta$ is the two photon detuning simulating the Zeeman field, the optical lattice potential $V(y)\propto |E_y|^2\cos^2(k_0y)$ comes from the AC stark shift by the standing wave, and the Raman potential $M_R(y)\propto E_{R,z}E_y\cos(k_0y)e^{iK_y y}$ is given by the two photon process formed by the standing waves and the Raman beam. The non-Hermiticity is in the damping of the spin down component, {\it i.e.} the $\gamma_\downarrow$ term, which could be achieved effectively by the transition between the spin down state and an excited atomic state with a short lifetime~\cite{Jo2022SM}, driven by the green beam in Fig.~\ref{Fig.1Dspec}(d). Under the s-band tight-binding approximation, the effective model is captured by
	\begin{align}
	H_{1D}	=& \sum_{j}(m_{z}+i\gamma_{\downarrow}/2)\bigr(|j\uparrow\rangle\langle j\uparrow|-|j\downarrow\rangle\langle j\downarrow|\bigr)-t_{0}\bigr(|j\uparrow\rangle\langle j+1\uparrow|-|j\downarrow\rangle\langle j+1\downarrow|+h.c.\bigr)\nonumber\\
	&+t_{so}\Big[e^{-iKj}\bigr(|j\downarrow\rangle\langle j+1\uparrow|-|j+1\downarrow\rangle\langle j,\uparrow|\bigr)+h.c.\Big],\label{eq.1Dmodel2}
	\end{align}
where a staggered gauge transformation on the spin down states $|j\downarrow\rangle\rightarrow (-1)^j |j\downarrow\rangle$ and an overall energy shift $-i\gamma_\downarrow/2$ have been adopted. And the coefficients are determined by
\begin{align}
	&t_0=\int dy \phi_{s}(y)[\frac{\hbar^2\partial_y^2}{2m}-V(y)]\phi_{s}(y-a),\nonumber\\
	&t_{so}=\int dy \phi_{s}(y)M_R(y)\phi_{s}(y-a),\ m_z=\delta.
	\end{align}
With a gauge transformation $|j\downarrow\rangle\rightarrow e^{-iKj}|j\downarrow\rangle$ we can rewrite the Hamiltonian just as Eq.~\eqref{eq.1Dmodel}.

	\begin{figure}[h]
	\centering
		\includegraphics[width=1.0\columnwidth]{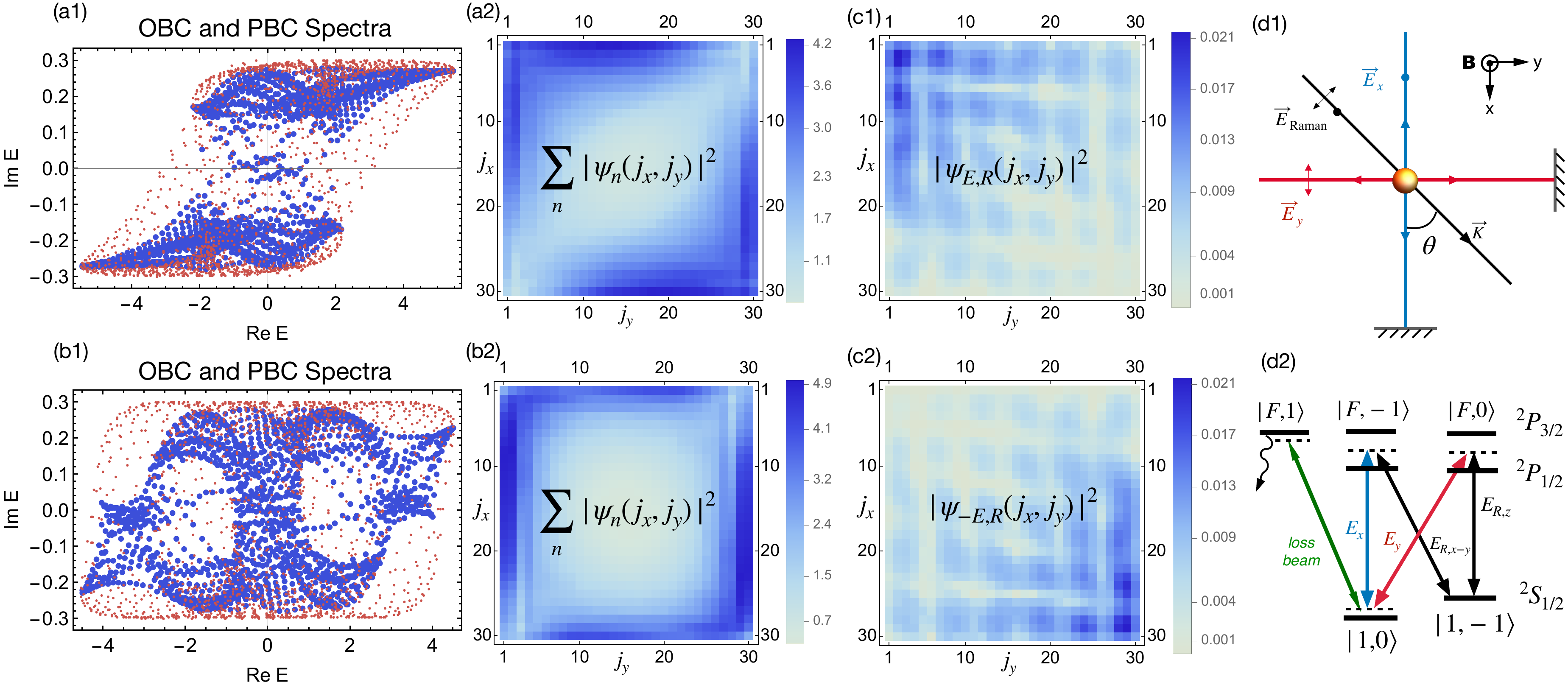}
		\caption{\label{Fig.2Dspec}Spectrum, wavefunction distribution and experiment scheme of the 2D models.  (a1, a2) The OBC (blue) and PBC (red) spectrums and the wavefunction distribution (summed over all eigenstates and spin) of the model in Eq.~\eqref{eq.2Dmodel} when $t_x=t_y=1, t_{so}^y=-it_{so}^x=t_{so}=0.4, m_z=1.3,\gamma_\downarrow=0.6,\vec{K}=\pi/a(\cos\theta,\sin\theta), \theta=50^\circ$ and (b1, b2) The results of Eq.~\eqref{eq.Dirac} when $t_x=t_y=1, t_{so}^y=t_{so}=0.4, m_z=0.3,\gamma_\downarrow=0.6,\theta=50^\circ$. (c1, c2) The right eigenvectors' distribution of a pair of skin modes of Eq.~\eqref{eq.2Dmodel} with eigenenergies $\pm E=\mp(1.63+0.17 \rm{I})$. (d1) The scheme of Raman optical lattice experiment to realize the 2D model in Eq.\eqref{eq.2Dmodel} (when $t_{so}^y=-it_{so}^x$). The $^{\rm{87}}\rm{Rb}$ atoms are confined in the 2D optical lattice formed by the standing waves (red and blue beams) and the spin states are defined to be the atomic states $|\uparrow\rangle=|1,-1\rangle,|\downarrow\rangle=|1,0\rangle$ in the $F=1$ manifold. (d2) All relevant transitions for the atoms, including the two photon process driven by the Raman beam (black running wave in (c)) and the standing waves, inducing spin-flipped couplings (SOC). The loss beam (green) is applied to drive the transition between the spin down state and excited states to give the non-Hermiticity just as in the 1D model.}
		\end{figure}

	\subsection{2D non-Hermitian models with PHS}
\begin{figure}[h]
	\centering
		\includegraphics[width=1.0\columnwidth]{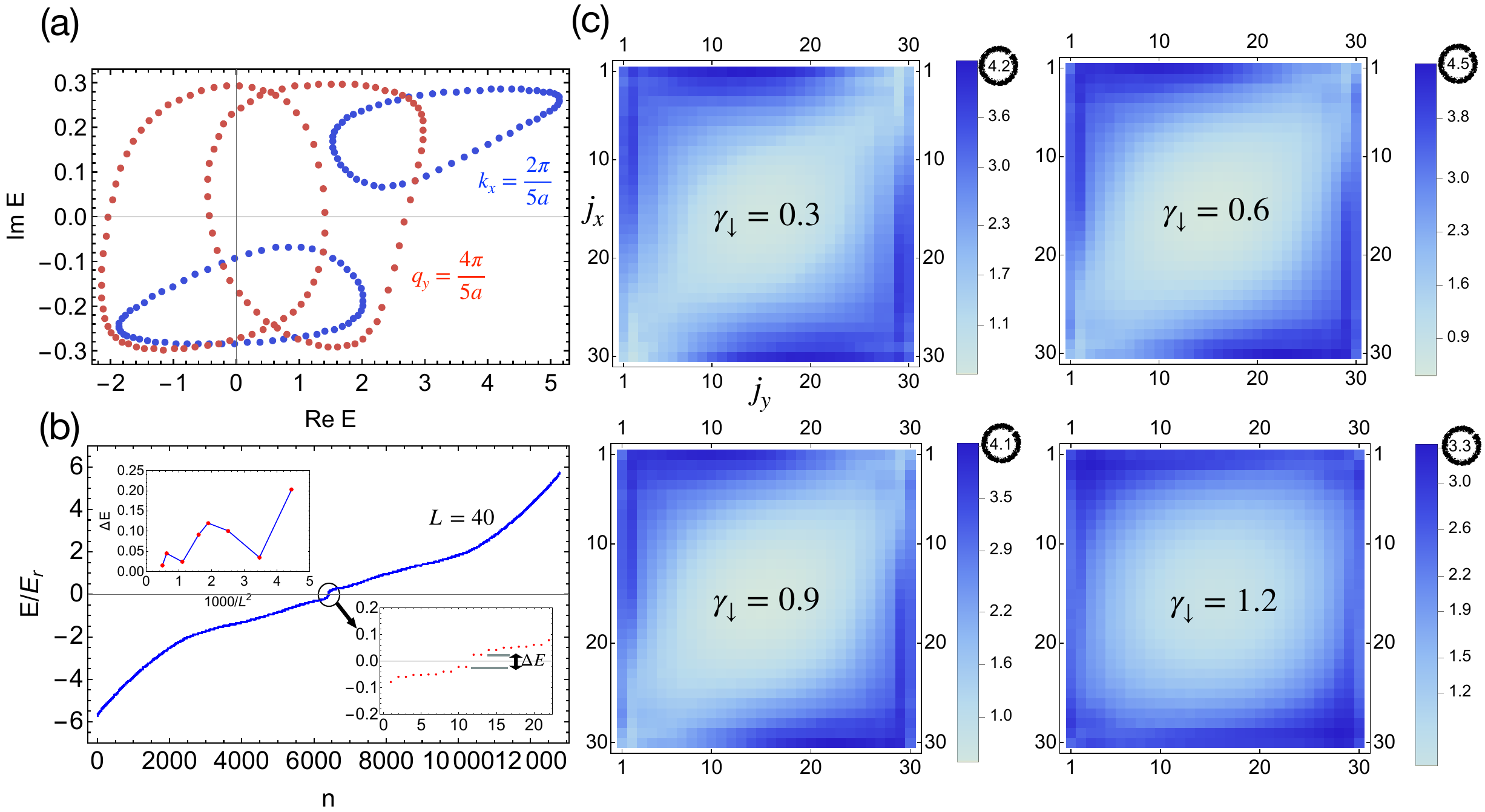}
		\caption{\label{Fig.varyingloss} The symmetric NHSE with varying loss rate $\gamma_\downarrow$. The localization of the skin modes, which is characterized by the maximum of the wavefunction distribution on the lattice (circled in the figure), reaches its maximum when it's close to the coefficients of the Hermitian part $t_{0}^{x/y}=1.0,t_{so}^{x/y}=0.4$.}
		\end{figure}

	Based on the researches on 2D Raman optical lattices~\cite{Liu2018SM, LiuPan2018SM}, we extend the 1D model to a 2D non-Hermitian lattice model:
	\begin{align}
	H_{2D}	=& \sum_{\vec{j}}\Big\{(m_{z}+i\gamma_{\downarrow}/2)\bigr(|\vec{j}\uparrow\rangle\langle\vec{j}\uparrow|-|\vec{j}\downarrow\rangle\langle \vec{j}\downarrow|\bigr)
	-\sum_{k=x,y}\Big[t_0^k\bigr(|\vec{j}\uparrow\rangle\langle\vec{j}+\vec{e}_k\uparrow|-e^{-i\vec{K}\cdot \vec{e}_k}|\vec{j}\downarrow\rangle\langle\vec{j}+\vec{e}_k\downarrow|\bigr)\nonumber\\
	 &+t_{\rm so}^k\bigr(|\vec{j}\downarrow\rangle\langle\vec{j}+\vec{e}_k\uparrow|-e^{i\vec{K}\cdot \vec{e}_k}|\vec{j}+\vec{e}_k\downarrow\rangle\langle\vec{j}\uparrow|\bigr)+h.c.\Big]\Big\},\label{eq.2Dmodel}
	\end{align}
respecting the PHS $S=UP=e^{-i\vec{K}\cdot\vec{j}}\sigma_x P$, where $\vec{K}$ is the wave vector of the Raman beam and $t_{0(\rm so)}^k,k=(x,y)$ are the spin-conserved(flipped) hopping coefficients in different directions. This model hosts symmetric corner skin effect due to the PHS as shown in Fig.~\ref{Fig.2Dspec}(a1, a2, c1, c2). Similar to the 1D case, to see the experimental feasibility we apply a gauge transformation $|\vec{j}\downarrow\rangle\rightarrow e^{-i\vec{K}\cdot\vec{j}}|\vec{j},\downarrow\rangle$ and rewrite the Hamiltonian as:
	\begin{align}
	H_{2D}	=& \sum_{\vec{j}}\Big\{(m_{z}+i\gamma_{\downarrow}/2)\bigr(|\vec{j}\uparrow\rangle\langle\vec{j}\uparrow|-|\vec{j}\downarrow\rangle\langle \vec{j}\downarrow|\bigr)
	-\sum_{k=x,y}\Big[t_0^k\bigr(|\vec{j}\uparrow\rangle\langle\vec{j}+\vec{e}_k\uparrow|-|\vec{j}\downarrow\rangle\langle\vec{j}+\vec{e}_k\downarrow|\bigr)\nonumber\\
	 &+t_{\rm so}^ke^{-i\vec{K}\cdot \vec{j}}\bigr(|\vec{j}\downarrow\rangle\langle\vec{j}+\vec{e}_k\uparrow|-|\vec{j}+\vec{e}_k\downarrow\rangle\langle\vec{j}\uparrow|\bigr)+h.c.\Big]\Big\},\label{eq.2Dmodel2}
	\end{align}
which is the low energy effective model under the s-band tight-binding approximation of the cold-atom system shown in Fig.~\ref{Fig.2Dspec}(d1, d2) described by
\begin{align}
	H_{2D}'	= &\int d^2\vec{r} \Big[ \sum_{s=\uparrow,\downarrow} |\vec{r}s\rangle\bigr(-\frac{\hbar^2\vec{\nabla}^2}{2m}+V(\vec{r})+\frac{\delta}{2}(\sigma_z)_{ss}\bigr)\langle\vec{r}s|\nonumber\\
	&-i\gamma_{\downarrow}|\vec{r}\downarrow\rangle\langle\vec{r}\downarrow|+\bigr(M_R(\vec{r})|\vec{r}\uparrow\rangle\langle\vec{r}\downarrow|+h.c.\bigr)\Big].
	\label{eq.2Dcontinous}
\end{align}
The optical lattice potential $V(\vec{r})= V_{0x}\cos^2(k_0x)+V_{0y}\cos^2(k_0y)$ (with $V_{0x/y}\propto |E_{x/y}^2|$) and Raman potential $M_R(\vec{r})=(M_{0x}\cos(k_0x)+M_{0y}\cos(k_0y))e^{i\vec{K}\cdot\vec{r}}$ (with$(M_{0x},M_{0y})\propto (E_{R}E_x,E_y E_{R})$) are induced by the standing waves and the Raman beam. The damping term is achieved by transitions to excited states just as in the 1D model case. The coefficients are determined by:
 \begin{align}
	&t_0^k=\int d^2\vec{r} \phi_{s}(\vec{r})[\frac{\hbar^2\vec{\nabla}^2}{2m}-V(\vec{r})]\phi_{s}(\vec{r}-a\vec{e}_k),\nonumber\\
	&t_{so}^k=\int d^2\vec{r} \phi_{s}(\vec{r})M_R(\vec{r})\phi_{s}(\vec{r}-a\vec{e}_k),\ m_z=\delta,
\end{align}
where $\psi_s(\vec{r})$ is the s-band Wannier function, so that the hopping coefficients could be controlled by tuning the beams independently. For example if we want $t_{so}^x=-it_{so}^y$ (so that a 2D Chern insulator is recovered without the loss term) the Raman beam could be circular polarized and the two linear components of it couples with the two optical beams respectively, driving the spin-flipped hoppings in the two directions in Eq.~\eqref{eq.2Dmodel2}. For the best observation of the symmetric NHSE, the effective loss rate $\gamma_\downarrow$ should be neither too small or large compared with $t_{0,so}^{x/y}$ because the system recovers the Hermiticity in the two extremes and the localization reaches it maximum in the middle regime, which is numerically illustrated by Fig.~\ref{Fig.varyingloss}. According to the previous studies~\cite{LiuPan2018SM, Jo2022SM} and the calculation with tight-binding approximation, here the appropriate parameters can be taken that $V_{0x/y}= 4.0,M_{0x/y}= 1.0,\delta=0.1,\gamma_\downarrow=0.1$ in the unit of the recoil energy $E_r=\frac{\hbar^2k_r^2}{2m}$, which give $t_{0}\approx 0.17,t_{so}^{x/y}\approx 0.07$ and $m_z=0.05$. The numerical results of the model with similar parameters given in~\ref{Fig.2Dspec} and~\ref{Fig.varyingloss} show the clear existence of NHSE in this regime.

	If the Raman beam's electric field is linearly polarized along $z$ direction, we realize a 2D Dirac semimetal with non-Hermiticity described by:
	\begin{align}
	H_{Dirac}	=& \sum_{\vec{j}}\Big\{(m_{z}+i\gamma_{\downarrow}/2)\bigr(|\vec{j}\uparrow\rangle\langle\vec{j}\uparrow|-|\vec{j}\downarrow\rangle\langle \vec{j}\downarrow|\bigr)
	-\sum_{k=x,y}\Big[t_0^k\bigr(|\vec{j}\uparrow\rangle\langle\vec{j}+\vec{e}_k\uparrow|-e^{-i\vec{K}\cdot \vec{e}_k}|\vec{j}\downarrow\rangle\langle\vec{j}+\vec{e}_k\downarrow|\bigr)\Big]\nonumber\\
	 &+t_{\rm so}\bigr(|\vec{j}\downarrow\rangle\langle\vec{j}+\vec{e}_y\uparrow|-e^{i\vec{K}\cdot \vec{e}_y}|\vec{j}+\vec{e}_y\downarrow\rangle\langle\vec{j}\uparrow|\bigr)+h.c.\Big]\Big\},\label{eq.Dirac}
	\end{align}
which respects the PHS $S=U\hat{P}=e^{-i\vec{K}\cdot\vec{j}}\sigma_x\hat{P}$ and hosts the symmetric NHSE are shown in Fig.~\ref{Fig.2Dspec}(b1, b2).

		\begin{figure}[h]
	\centering
		\includegraphics[width=0.7\columnwidth]{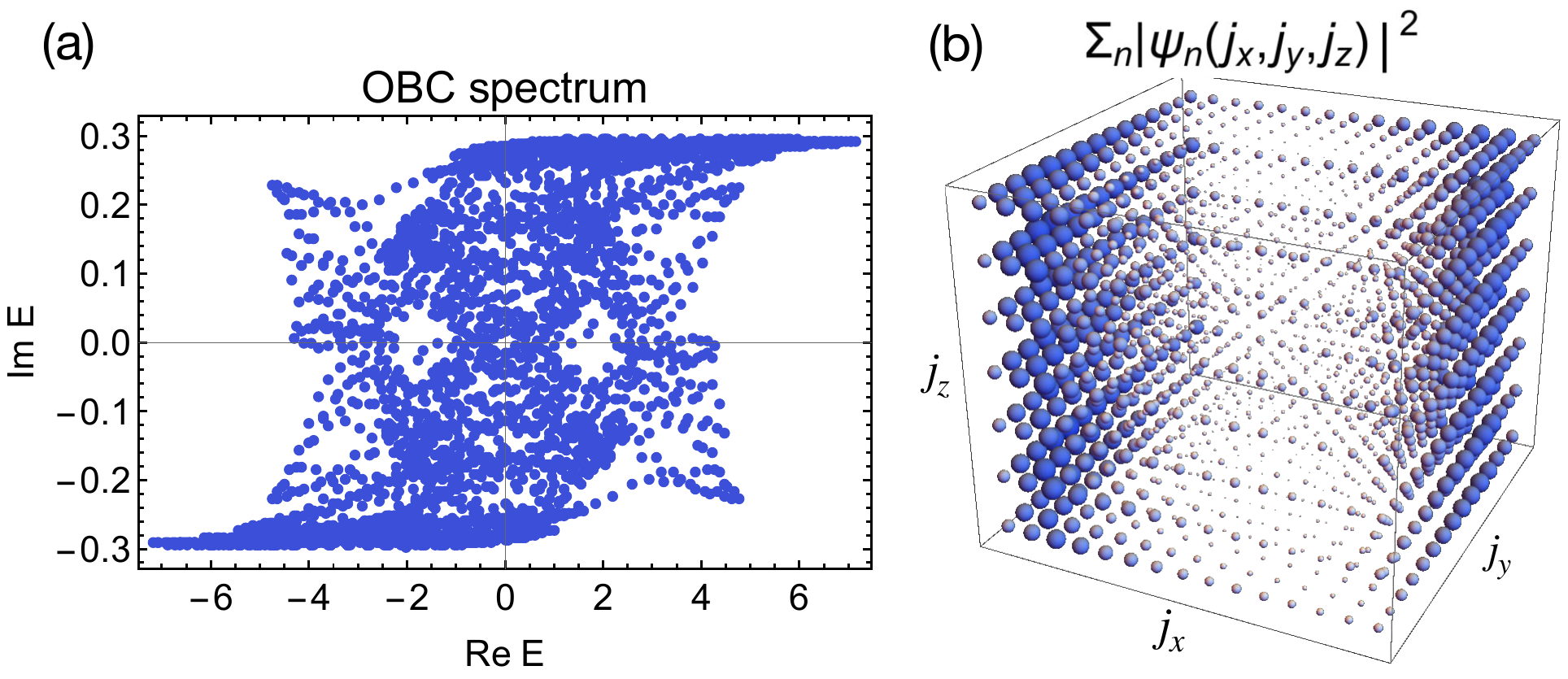}
		\caption{\label{Fig.3D} OBC spectrum and wavefunction distribution of the 3D non-Hermitian Weyl semimetal in Eq.~\eqref{eq.3D} where $t_x=t_y=t_z=1, t_{so}=0.4, m_z=0.3,\gamma_\downarrow=0.6,\vec{K}=\pi/a(\cos\theta,\sin\theta,0), \theta=50^\circ$ with the size of the system $L_x=L_y=15,L_z=7$.  (a) The OBC spectrum on the complex plane. (b) The wavefunction distribution on the 3D lattice summed over all eigenstates and spin degree of freedom, in which the bigger blue dots mean larger value than the white small dots.}
		\end{figure}

	\subsection{3D non-Hermitian Weyl semimetal with PHS}	
A 3D non-Hermitian Weyl semimetal lattice model is given by
	\begin{align}
	H_{3D}	=& \sum_{\vec{j}}\Big\{(m_{z}+i\gamma_{\downarrow}/2)\bigr(|\vec{j}\uparrow\rangle\langle\vec{j}\uparrow|-|\vec{j}\downarrow\rangle\langle \vec{j}\downarrow|\bigr)
	-\sum_{k=x,y,z}\Big[t_0^k\bigr(|\vec{j}\uparrow\rangle\langle\vec{j}+\vec{e}_k\uparrow|-e^{-i\vec{K}\cdot \vec{e}_k}|\vec{j}\downarrow\rangle\langle\vec{j}+\vec{e}_k\downarrow|\bigr)\Big]\nonumber\\
	 &+\sum_{k=x,y}t_{\rm so}^k\bigr(|\vec{j}\downarrow\rangle\langle\vec{j}+\vec{e}_k\uparrow|-e^{i\vec{K}\cdot \vec{e}_k}|\vec{j}+\vec{e}_k\downarrow\rangle\langle\vec{j}\uparrow|\bigr)+h.c.\Big]\Big\},\label{eq.3D}
	\end{align}
whose OBC spectrum and symmetric skin effect are shown in Fig.~\ref{Fig.3D}. This model could be realized by replacing the trap along $z$ direction by an optical lattice beam in the 2D setup of Fig.~\ref{Fig.2Dspec}(d1), and the frequency of this standing wave has a detuning from the $x,y$ directions' optical beams to prevent the spin-flipped hoppings along $z$~\cite{Liu2020SM, LiuPan2021SM}.

\end{document}